\documentclass{cmslatex}
\usepackage{latexsym, amssymb, enumerate, amsmath}

\renewcommand{\theequation}{\arabic{section}.\arabic{equation}}

\usepackage{dsfont}
\usepackage{lineno}
\usepackage{subfigure}
\usepackage{graphicx}
\usepackage{multirow}
\usepackage{color}
\usepackage{xspace}
\usepackage{pdfsync}
\usepackage{hyperref} 
\usepackage{algorithmicx}
\usepackage{algpseudocode}
\usepackage{algorithm}
\graphicspath{{../Figures/}}

\newcommand{\bq}{\begin{equation}}
\newcommand{\eq}{\end{equation}}
\newcommand{\R}{\mathbb{R}}
\newcommand{\E}{\mathbb{E}}
\newcommand{\abs}[1]{\left\vert#1\right\vert}
\newcommand{\norm}[1]{\left\vert#1\right\vert}

\newcommand{\bO}{\mathcal{O}}

\newcommand{\Dt}{\mathcal{D}}

\newcommand{\M}{\mathcal{M}}

\newcommand{\MA}{Monge-Amp\`ere\xspace}
\newcommand{\vp}{v^\perp}

\algnewcommand{\LineComment}[1]{\State \(\triangleright\) #1}

\begin{document}

\title{Application of the {W}asserstein metric to seismic signals
\thanks{\today
}}
\author{Bj\"orn Engquist
\thanks {Department of Mathematics and ICES, The University of Texas at Austin, 1 University Station C1200, Austin, TX 78712 USA, (engquist@math.utexas.edu).  \newline The first author was partially supported by NSF DMS-1217203.}
\and Brittany D. Froese \thanks {Department of Mathematics and ICES, The University of Texas at Austin, 1 University Station C1200, Austin, TX 78712 USA, (bfroese@math.utexas.edu). \newline The second author was partially supported by an NSERC PDF.}}



\pagestyle{myheadings} \markboth{Application of the {W}asserstein metric to seismic signals}{Bj\"orn Engquist and Brittany D. Froese}\maketitle

\begin{abstract}
Seismic signals are typically compared using travel time difference or $L_2$ difference.  We propose the Wasserstein metric as an alternative measure of fidelity or misfit in seismology.  It exhibits properties from both of the traditional measures mentioned above.   The numerical computation is based on the recent development of fast numerical methods for the \MA equation and optimal transport. Applications to waveform inversion and registration are discussed and simple numerical examples are presented. 
\end{abstract}


\section{Introduction}
A classical way of comparing seismic signals is to use the travel time difference.  This can be done in the time domain or, more recently, by different measures of the phase shift; see for example~\cite{vanLeeuwen}.
The misfit or fidelity estimate can be between two measured signals or between a computed signal and a measured seismic signal.  For more complex signals, travel time estimates may not be appropriate and $L_2$ estimates of misfit are often used in full waveform inversion~\cite{PrattSeismic,TarantolaInversion}.

We propose the Wasserstein metric as a measure of misfit that combines many of the best features of the metrics given by travel time and $L_2$.  The Wasserstein metric measures the difference between two distributions by the optimal cost of rearranging one distribution into the other~\cite{Villani}.  The mathematical definition of the distance between the distributions $f:X\to\R$, $g:Y\to\R$ can be formulated as
\bq\label{eq:W2}  W_2^2(f,g) = \inf\limits_{T\in\M} \int\limits_X\abs{x-T(x)}^2f(x)\,dx \eq
where $\M$ is the set of all maps that rearrange the distribution $f$ into $g$.

We consider two simple one-dimensional examples to show the relation of the Wasserstein metric to travel time and $L_2$.  First, we compare the hat functions $f(x)$ and $g(x,s) = f(x-s)$ with 
\bq\label{eq:hat}  f(x) = \max\{1-\abs{x},0\}.\eq
For small $s$, the $L_2$ distance is 
\[ \|f-g\|_{L_2}^2 = 2s^2 + \bO(s^3), \]
and the Wasserstein metric measures the misfit by
\[ W_2^2(f,g) = s^2. \]

For large $s$, on the other hand, the $L_2$ distance is given by
\[ \|f-g\|_{L_2}^2 = \|f\|_{L_2}^2 + \|g\|_{L_2}^2 = 2, \]
which is independent of $s$ and of no value in minimisation. However, the Wasserstein distance preserves the ideal $\bO(s^2)$ scaling.

\begin{figure}[htdp]
\begin{center}
\subfigure[]{\includegraphics[width=0.6\textwidth]{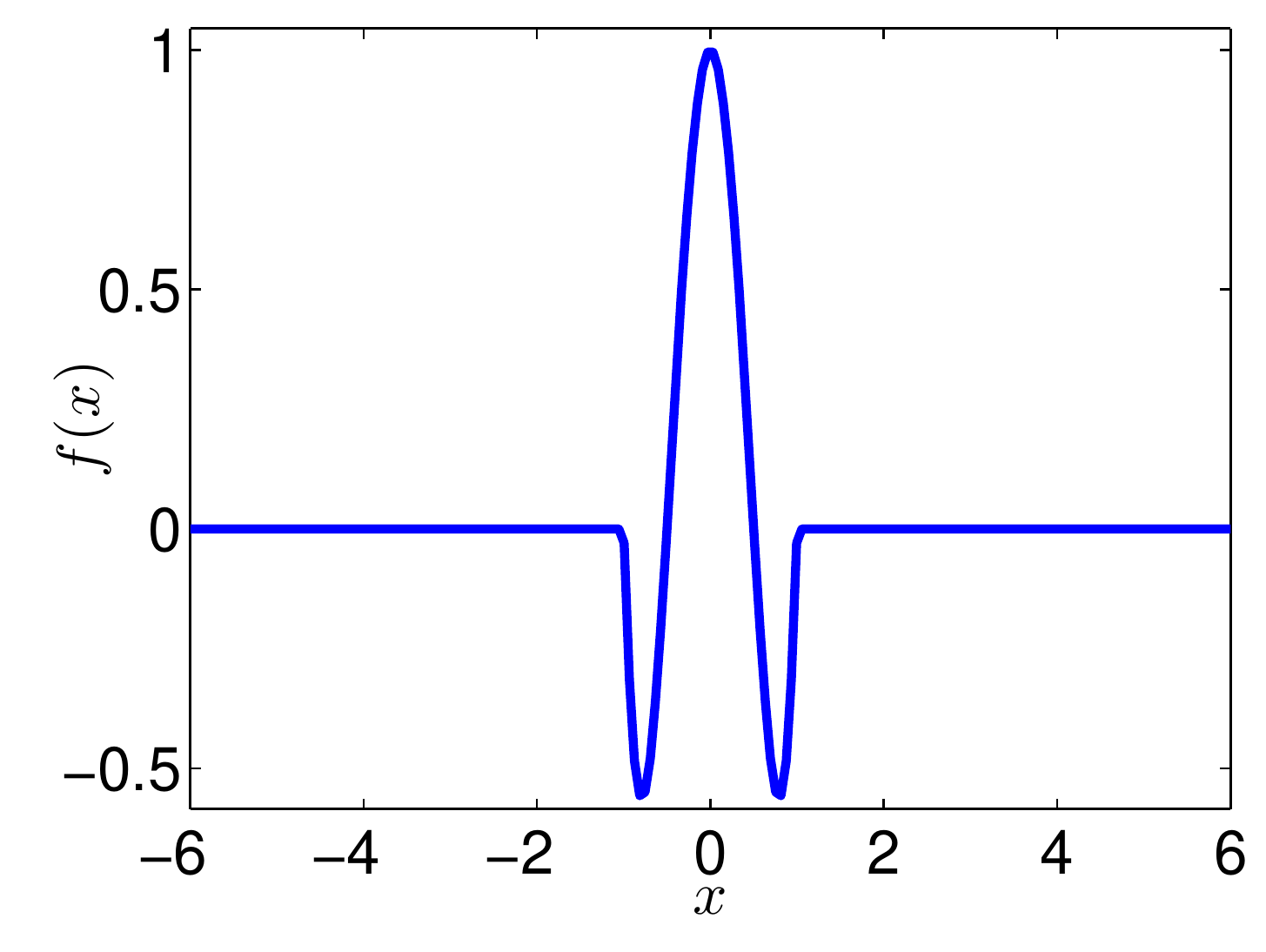}\label{fig:wavelet_prof}}
\subfigure[]{\includegraphics[width=0.6\textwidth]{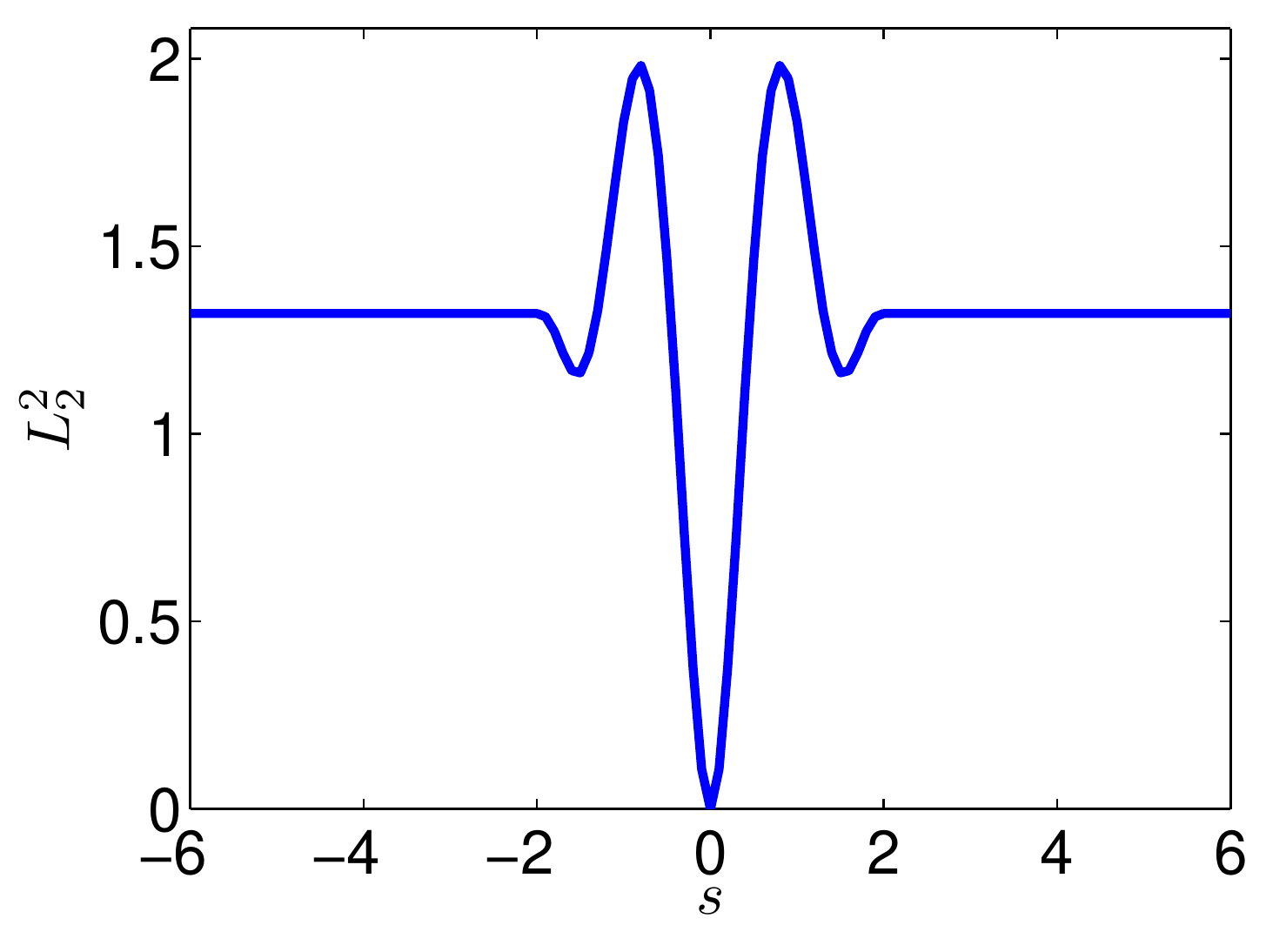}\label{fig:wavelet1d_L2}}
\subfigure[]{\includegraphics[width=0.6\textwidth]{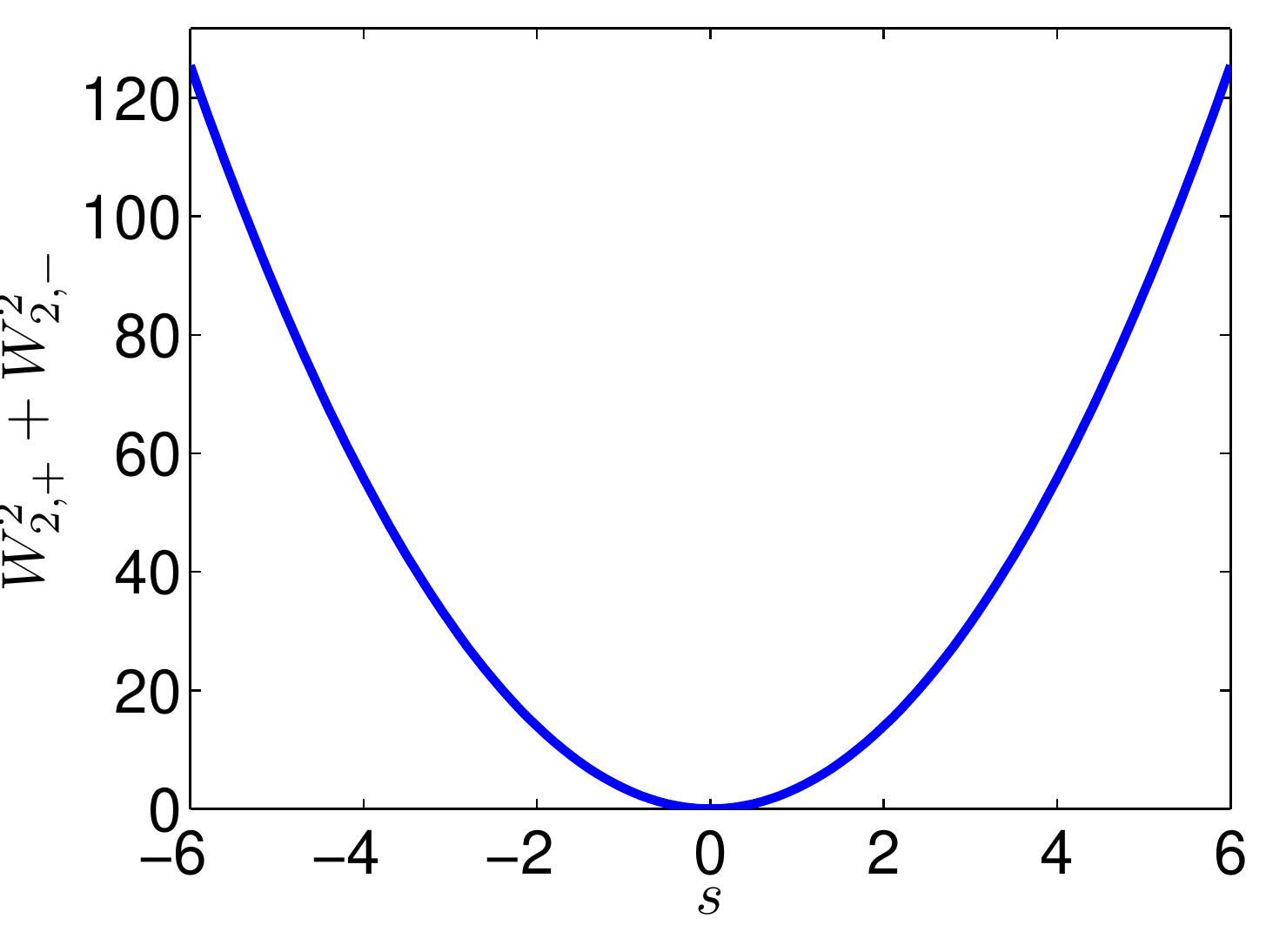}\label{fig:wavelet1d_W2_pm}}
\caption{\subref{fig:wavelet_prof}~A wavelet profile $f(x)$.  The distances between $f(x)$ and $g(x) = f(x-s)$ measured by \subref{fig:wavelet1d_L2}~$L_2^2(f,g)$, and \subref{fig:wavelet1d_W2_pm}~$W_2^2(f^+,g^+) + W_2^2(f^-,g^-)$. }
\label{fig:wavelet}
\end{center}
\end{figure}

For an example that better resembles seismic signals, consider the misfit between the simple wavelet $f$ in Figure~\ref{fig:wavelet_prof} and another wavelet shifted by a distance $s$.  Figures~\ref{fig:wavelet1d_L2}-\ref{fig:wavelet1d_W2_pm} clearly illustrate the advantage of using the Wasserstein distance in minimisation processes.

Earlier algorithms for the numerical computation of the Wasserstein distance required a large number of operations~\cite{Bertsekas,Bosc, BenBre}.  The optimal transportation problem can be rigorously related to the following \MA equation~\cite{Brenier, KnottSmith}, which enables the construction of more efficient methods for computing the Wasserstein distance.
\bq\label{eq:MA} \begin{cases}
\det(D^2u(x)) = f(x)/g(\nabla u(x)) + \langle u \rangle, & x\in X\\
\nabla u(X) = Y\\
u \text{ is convex.}
\end{cases}\eq
The Wasserstein distance is then given by
\bq\label{eq:W2_u}
W_2^2(f,g) = \int\limits_X \abs{x-\nabla u(x)}^2f(x)\,dx.
\eq
There are now fast and robust numerical algorithms for the solution of~\eqref{eq:MA}, and thus for the computation of $W_2^2$~\cite{BFO_OTNum}.

The solution $u$ of the \MA equation contains additional information since the vector $\nabla u(x) - x$ indicates which parts of the distributions $f$ and $g$ are connected under the optimal transport map.   This information is useful for problems in image registration~\cite{HakerRegistration}, meteorology~\cite{Cullen}, mesh generation~\cite{Budd}, reflector design~\cite{GlimmOlikerReflectorDesign}, and astrophysics~\cite{FrischUniv}.  As we will see, it can also be of great value in seismology.

\section{Challenges in application to seismology}
While the Wasserstein distance has many excellent properties, there remain challenges that need to be addressed due to the specific nature of seismic signals.  Some of these difficulties come from the formulation of the Wasserstein metric and some from the numerical algorithm used to solve the \MA equation.

\subsection{Positivity}
The Wasserstein metric requires $f,g \geq 0$, which is typically not the case with seismic signals.  This could be achieved by adding positive constants to $f$ and $g$, but this would distort the optimal transportation map.  Another option is to compare envelopes of the functions.  We have chosen to compare separately the positive and negative parts of $f = f^+-f^-$, $g = g+-g^-$, then add the results.  Then the misfit we compute is
\[ W_2^2(f^+,g^+) + W_2^2(f^-,g^-). \]

\subsection{Mass conservation}
The Wasserstein metric also requires that mass is conserved,
\[ \int\limits_Xf(x)\,dx = \int\limits_Yg(y)\,dy. \]
This can be achieved most easily by a simple constant scaling of the densities $f$ and $g$.  

Other scalings are also possible when additional information about the distribution is available.  For instance, if the signals $f$ and $g$ are close to each other, with each consisting of two separate components, we could rescale the components separately.  This would reduce the risk that the optimal transportation plan $T(x)$ could transport some mass between separate components, which may not be desirable in certain applications.

\subsection{Convexity}
Other challenges come from the solution of the \MA equation.  In particular, the set $Y$ where the target density $g(y)$ is positive must be convex.  This can be accomplished by first selecting a convex set $\tilde{Y}$ containing $Y$, then preprocessing the data as follows,
\bq \tilde{g}^+(y) = \begin{cases}
g^+(y) + \theta, &y \in \tilde{Y}^+\\
0, & \text{otherwise}.
\end{cases}\eq
Here $\theta>0$, $\tilde{Y}^+$ is a convex set containing the support of $g^+$, and the same type of transformation can be applied to the other components $f^+, f^-, g^-$.  In the examples below, we choose the convex sets to be rectangles.  This $\theta$-layer will introduce a small amount of artificial transport into and out of the layer.  To reduce distortion of the optimal transportation plan, we will choose all rectangles to have the same size, with each one centred at the centre of mass of the corresponding distribution.  If the optimal transportation plan itself is of interest, we can reduce some of the artificial transport by thresholding the transport vectors $\nabla u(x)-x$ to zero in the layer.

\subsection{Regularity}
The numerical method used to solve the \MA equation also requires that the ratio $f(x)/g(y)$ is Lipschitz continuous in the $y$ variable.  This leads to the requirement that $\theta$ cannot be too small, particularly if there are regions where $g=0$ and the corresponding gradient is not small.  Appropriate choice of $\theta$ and convolution of $g$ (and $f$ for symmetry) with a regularising kernel will ensure the success of the numerical method.

\section{Numerical algorithm}
We describe here a two-dimensional form of the algorithm we use to solve the \MA equation.  The method is based on the following variational characterisation of the \MA equation~\eqref{eq:MA} combined with the convexity constraint~\cite{FroeseTransport,FOMATheory}.
\bq\label{eq:MA_convex}
{\det}^+(D^2u) = \min\limits_{\{\nu_1,\nu_2\}\in V}\left\{\max\{u_{\nu_1,\nu_1},0\} \max\{u_{\nu_2,\nu_2},0\}+\min\{u_{\nu_1,\nu_1},0\} + \min\{u_{\nu_2,\nu_2},0\}\right\}
\eq
where $V$ is the set of all orthonormal bases for $\R^2$.  The transportation constraint $\nabla u(X) = Y$ can be re-expressed as the Hamilton-Jacobi equation
\bq\label{eq:HJ} H(\nabla u(x)) = 0, \quad x\in\partial X \eq
where $H(y)$ is the signed distance to the convex target set $Y$~\cite{BFO_OTNum}.

The discretisation described below leads to a large system of nonlinear equations, which is solved using Newton's method.  The linear equations arising in each Newton iteration are solved using a direct sparse solver.  In a typical example, fewer than ten iterations are required for convergence.

\subsection{Monotone approximation of the \MA equation}
Equation~\eqref{eq:MA_convex} can be discretised by computing the minimum over finitely many directions $\{\nu_1,\nu_2\}$, which may require the use of a wide stencil.  For simplicity and brevity, we describe a compact version of the scheme and refer to~\cite{FOMATheory} for complete details.

We begin by introducing the finite difference operators
\begin{align*}
[\Dt_{x_1x_1}u]_{ij} &= \frac{1}{dx^2} 
\left(
{u_{i+1,j}+u_{i-1,j}-2u_{i,j}}
\right)
\\
[\Dt_{x_2x_2}u]_{ij} &= \frac{1}{dx^2}
\left(
u_{i,j+1}+u_{i,j-1}-2u_{i,j}
\right)
\\
[\Dt_{x_1}u]_{ij} &= \frac{1}{2dx}
\left(
u_{i+1,j}-u_{i-1,j}
\right)\\
[\Dt_{x_2}u]_{ij} &= \frac{1}{2dx}
\left(
u_{i,j+1}-u_{i,j-1}
\right)\\
[\Dt_{vv}u]_{ij} &= \frac{1}{2dx^2}\left(u_{i+1,j+1}+u_{i-1,j-1}-2u_{i,j}\right)\\
[\Dt_{\vp\vp}u]_{ij} &= \frac{1}{2dx^2}\left(u_{i+1,j-1}+u_{i+1,j-1}-2u_{i,j}\right)\\
[\Dt_{v}u]_{ij} &= \frac{1}{2\sqrt{2}dx}\left(u_{i+1,j+1}-u_{i-1,j-1}\right)\\
[\Dt_{\vp}u]_{ij} &= \frac{1}{2\sqrt{2}dx}\left(u_{i+1,j-1}-u_{i-1,j+1}\right).
\end{align*}

In the compact version of the scheme, the minimum in~\eqref{eq:MA_convex} is approximated using only two possible values.  The first uses directions aligning with the grid axes.
\begin{multline}\label{MA1}
MA_1[u] = \max\left\{\Dt_{x_1x_1}u,\delta\right\}\max\left\{\Dt_{x_2x_2}u,\delta\right\} \\- \min\left\{\Dt_{x_1x_1}u,\delta\right\} - \min\left\{\Dt_{x_2x_2}u,\delta\right\} - f / g\left(\Dt_{x_1}u, \Dt_{x_2}u\right) - u_0.
\end{multline}
Here $dx$ is the resolution of the grid, $\delta>K\Delta x/2$ is a small parameter that bounds second derivatives away from zero, $u_0$ is the solution value at a fixed point in the domain, and $K$ is the Lipschitz constant in the $y$-variable of $f(x)/g(y)$.

For the second value, we rotate the axes to align with the corner points in the stencil, which leads to
\begin{multline}\label{MA2}
MA_2[u] = \max\left\{\Dt_{vv}u,\delta\right\}\max\left\{\Dt_{\vp\vp}u,\delta\right\} - \min\left\{\Dt_{vv}u,\delta\right\} - \min\left\{\Dt_{\vp\vp}u,\delta\right\}\\ - f / g\left(\frac{1}{\sqrt{2}}(\Dt_{v}u+\Dt_{\vp}u), \frac{1}{\sqrt{2}}(\Dt_{v}u-\Dt_{\vp}u)\right) - u_0.
\end{multline}

Then the compact monotone approximation of the \MA equation is
\bq\label{eq:mA_compact} -\min\{MA_1[u],MA_2[u]\} = 0. \eq

\subsection{Monotone approximation of the Hamilton-Jacobi boundary condition}
We describe the boundary conditions in the special case where the source and target sets are rectangles and refer to~\cite{BFO_OTNum} for details of the more general setting.
In this setting, the Hamilton-Jacobi equation can be written as the Neumann boundary condition
\bq\label{eq:neumann}
u_{x_1}(x_1^{min}) = y_1^{min}, \, u_{x_1}(x_1^{max}) = y_1^{max}, \,u_{x_2}(x_2^{min}) = y_2^{min}, \,u_{x_2}(x_2^{min}) = y_2^{min},
\eq
which transports each side of the domain $X$ to the corresponding side of the target rectangle $Y$.
A monotone discretisation is easily constructed; on the left side of the domain, for example,
\bq\label{eq:leftBC} \frac{u_{2,j} - u_{1,j}}{dx} =  y_1^{min}.\eq 

\subsection{Filtered approximation}
For improved accuracy, we can combine the monotone scheme $F_M$ just described with a scheme that is formally more accurate $F_A$.  A second-order scheme is easily constructed using a centred difference discretisation of the two-dimensional \MA equation
\[ u_{xx}u_{yy} - u_{xy}^2 = f/g(u_x, u_y) \]
and a one-sided second-order approximation of the Neumann boundary conditions.
We introduce a filter such as
\bq\label{eq:filter}
S(x) = \begin{cases}
x & \norm{x} \leq 1 \\
0 & \norm{x} \ge 2\\
-x+ 2  & 1\le x \le 2 \\
-x-2  & -2\le x\le -1.
\end{cases} 
\eq
Then a convergent, higher-order scheme is given by
\bq\label{eq:filtered}
F_F[u] = F_M[u] + \epsilon S\left(\frac{F_A[u]-F_M[u]}{\epsilon}\right).
\eq
In many cases, the higher order accuracy can be achieved even when the filtered scheme is based on a compact monotone scheme; see~\cite{FOFiltered}.

\section{Numerical examples for a two-layer material}

We consider the example of the response from a two-layer material width upper and lower depths $d_1, d_2$ and wave speeds $v_1, v_2$ respectively; see Figure~\ref{fig:commonshot}.  Typical seismic signals are pictured in Figure~\ref{fig:source}-\ref{fig:target} in the offset-time domain.  After preprocessing, we compute the optimal transportation plan between these two distributions.  

\subsection{Inversion}
One potential application of optimal transportation is full waveform inversion.  To accomplish this, it is necessary to determine the unknown parameters that minimise the misfit between the observed and synthetic signals.  In this example, the unknown parameters are $d_1, d_2, v_1, v_2$.  To demonstrate the advantage of the Wasserstein metric as a measure of misfit, we fix one signal $g$ that is computed using $d_1 = 1, d_2 = 0.5, v_1 = 1, v_2 = 1.5$.  We then compute the distance $W_2^2(f,g)$, where $f$ is computed for a range of different parameter values.  We plot several cross-sections of this distance in Figure~\ref{fig:cross_sections}.  For comparison, we also plot several cross-sections of the $L_2^2$ distance $\|f-g\|_{L_2}^2$.  It is clear that the Wasserstein distance is much more suitable for minimisation.  In preliminary computations, the minimisation has been successfully accomplished using a simple Matlab implementation of the Nelder-Mead simplex method.

\subsection{Registration}
\begin{figure}[htdp]
\centering
\subfigure[]{\includegraphics[width=0.45\textwidth]{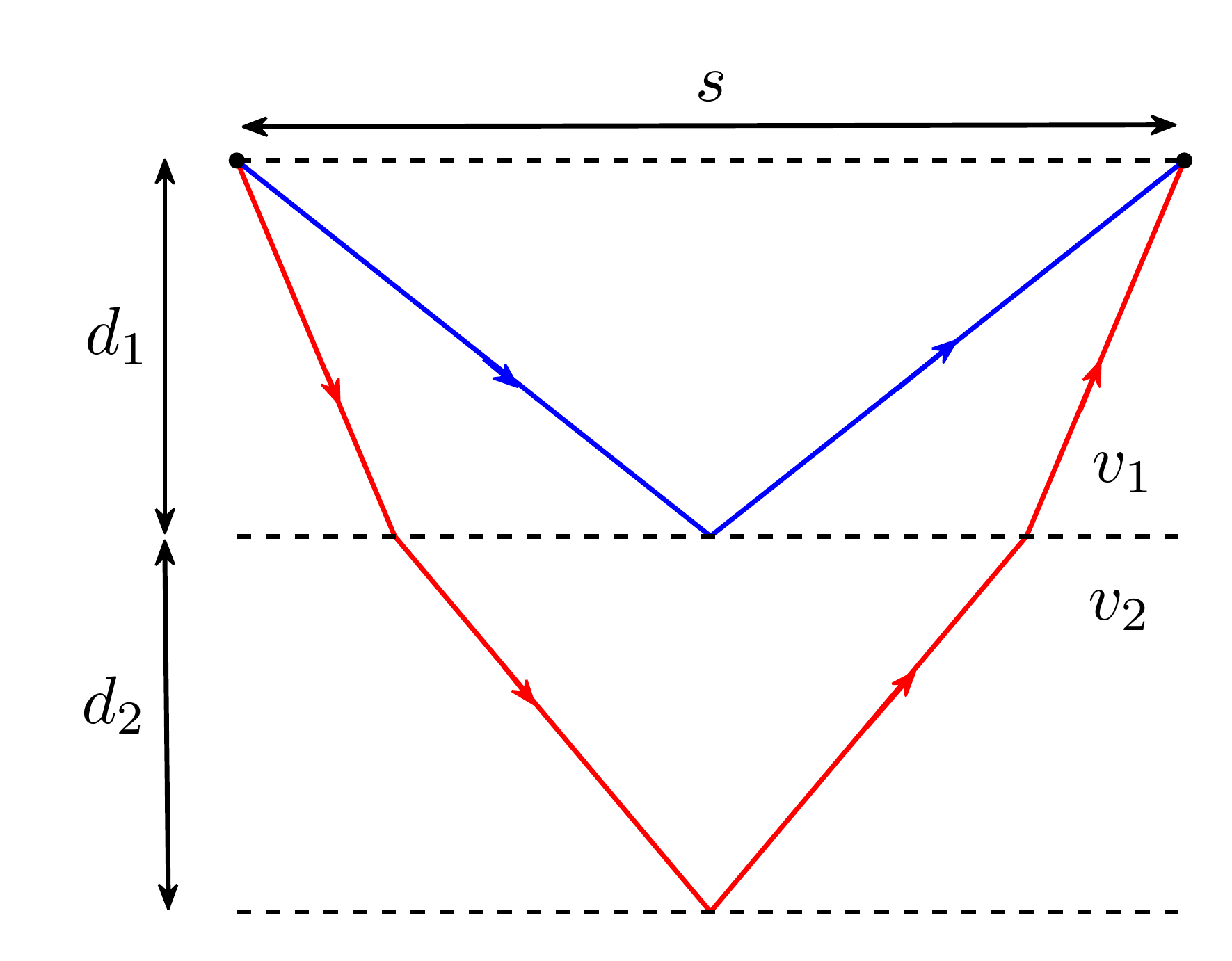}\label{fig:commonshot}}
{\subfigure[]{\includegraphics[width=0.45\textwidth]{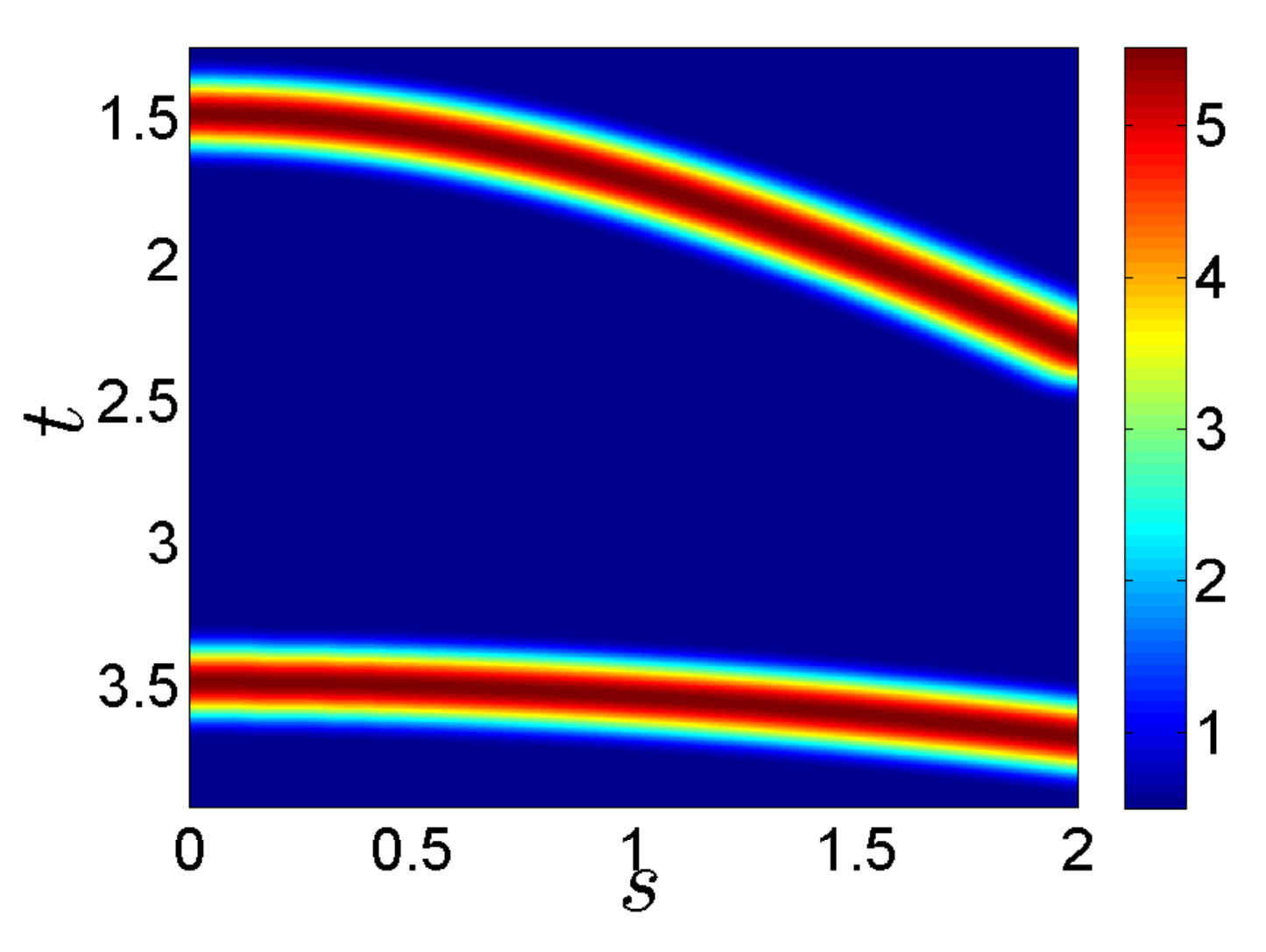}\label{fig:source}}}
{\subfigure[]{\includegraphics[width=0.45\textwidth]{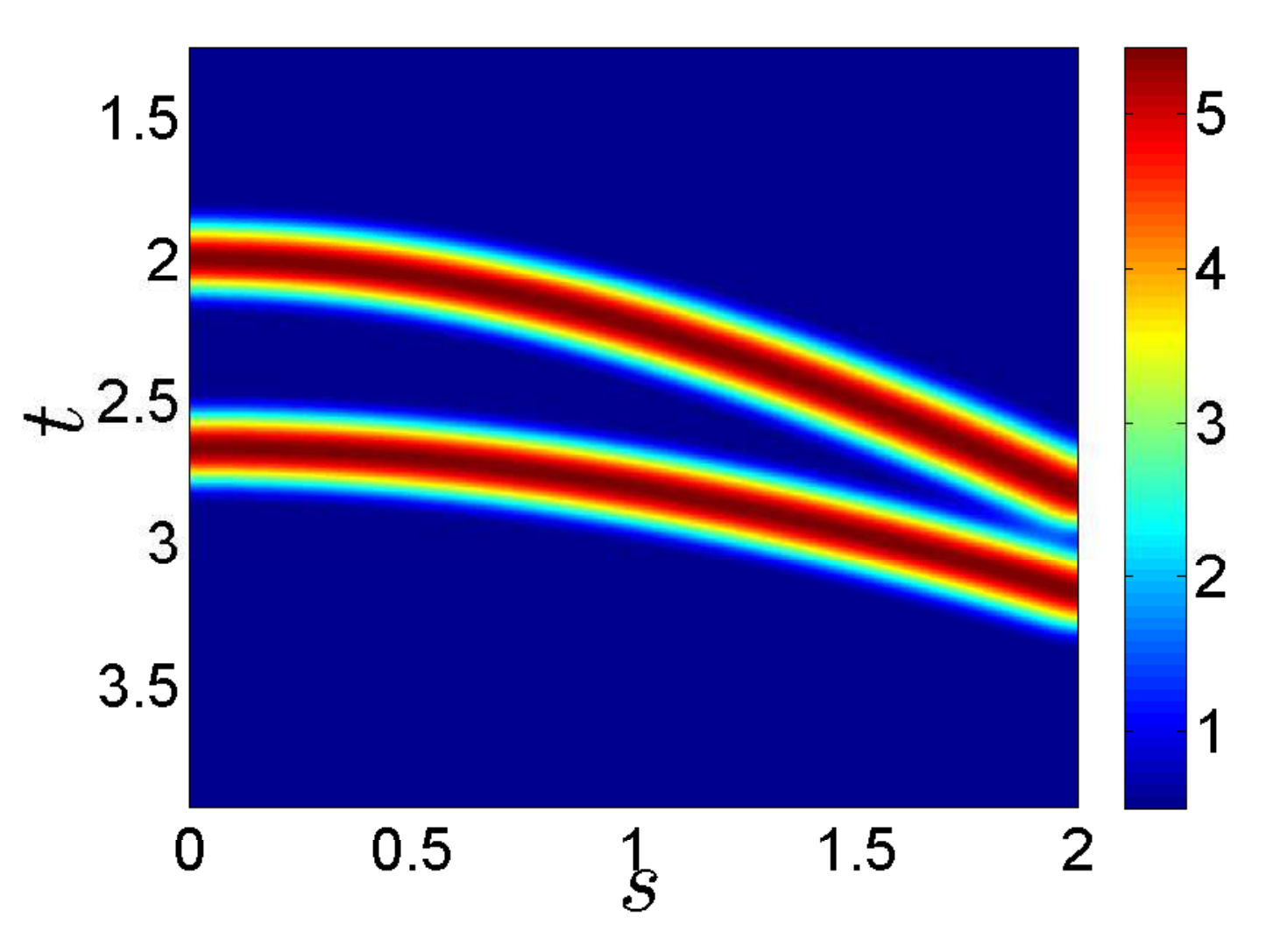}\label{fig:target}}}
{\subfigure[]{\includegraphics[width=0.45\textwidth]{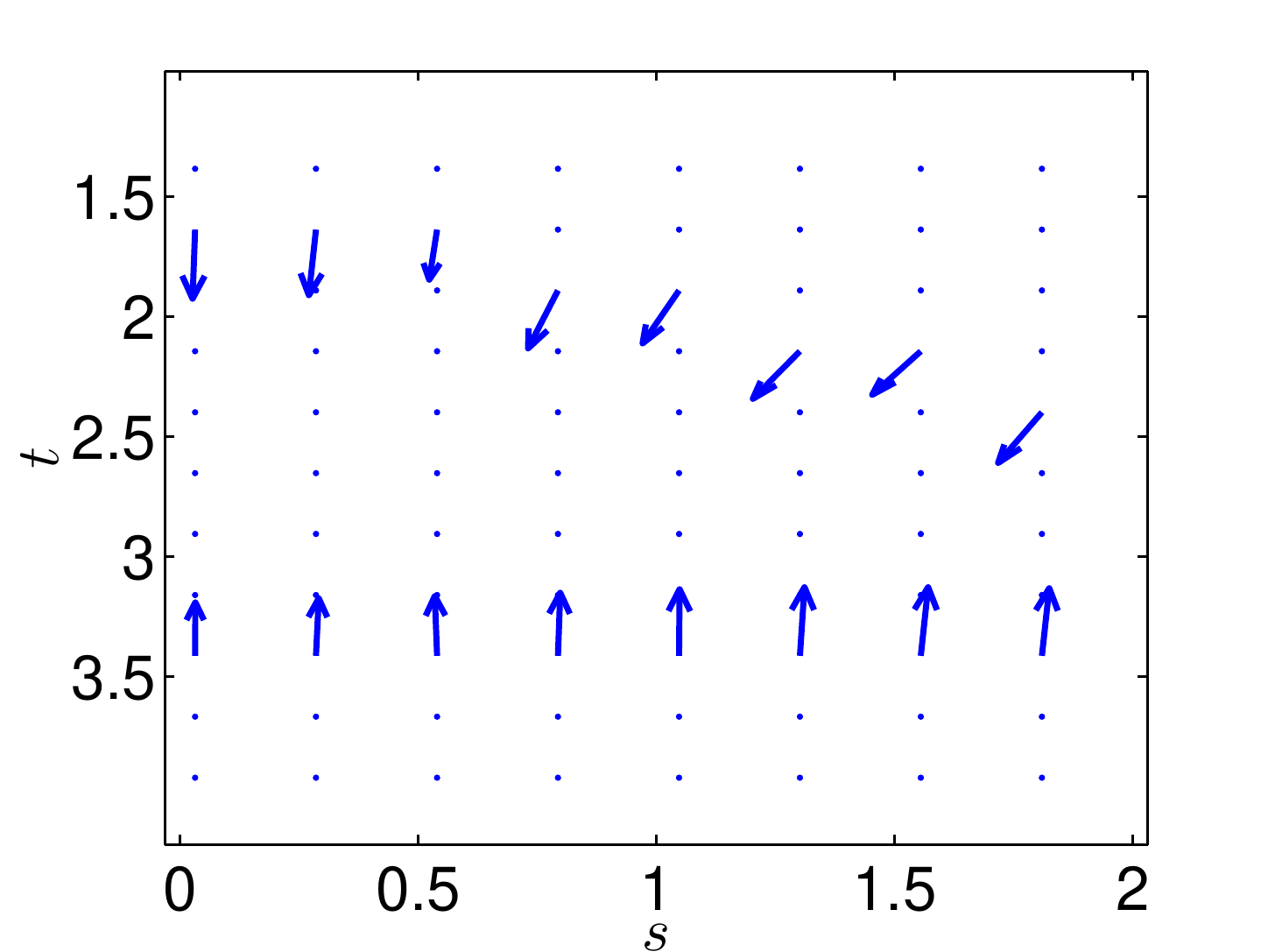}\label{fig:map}}}
\caption{\subref{fig:commonshot}~A two-layer material, \subref{fig:source},\subref{fig:target}~seismic signals from different materials, and \subref{fig:map}~the displacement $\nabla u (x) - x$ coming from the optimal transportation map that defines registration between the two signals. }
\label{fig:layer}
\end{figure}

\begin{figure}[htdp]
\begin{center}
\subfigure[]{\includegraphics[width=0.45\textwidth]{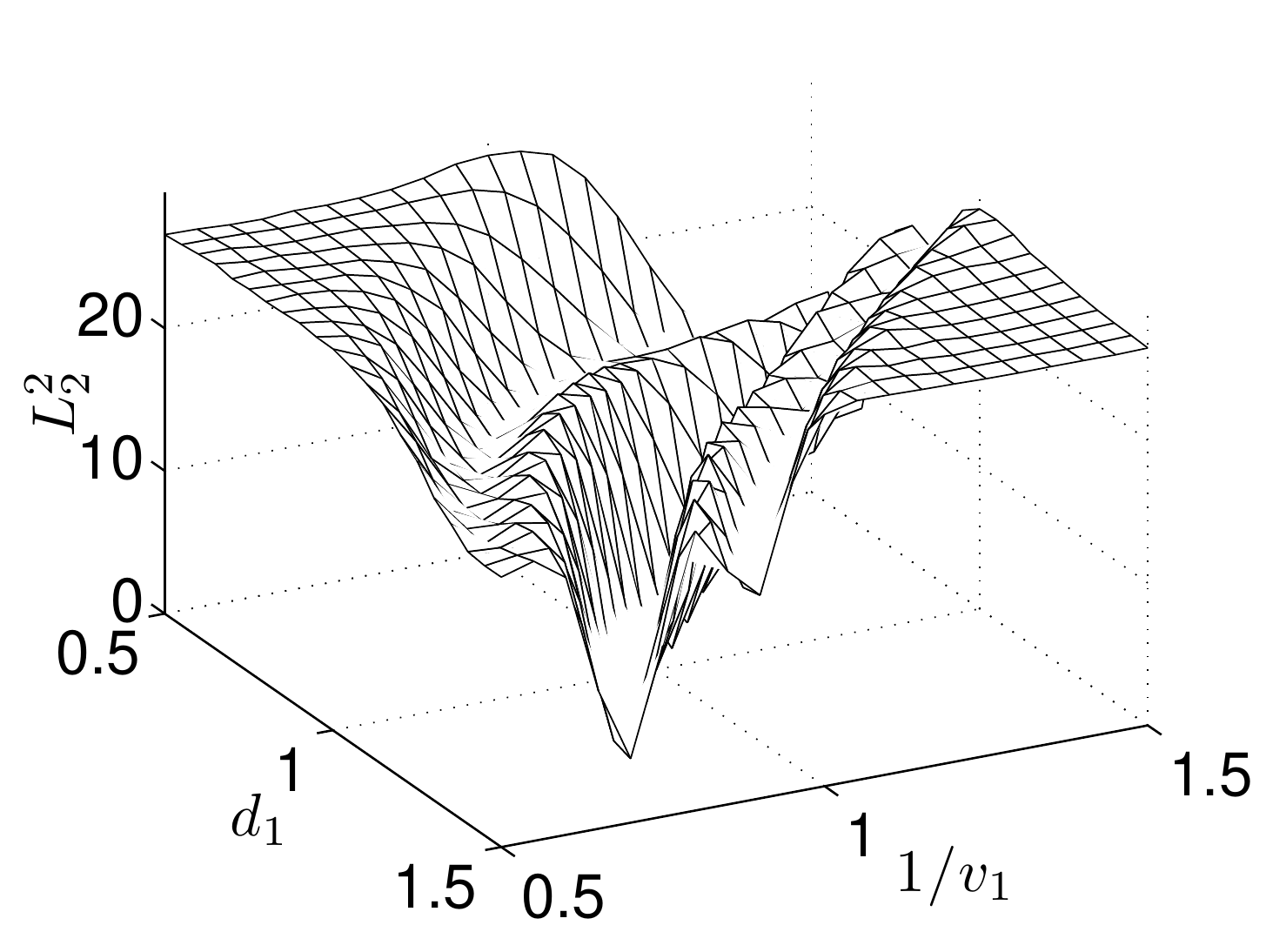}\label{fig:L2_d1v1_mesh}}
\subfigure[]{\includegraphics[width=0.45\textwidth]{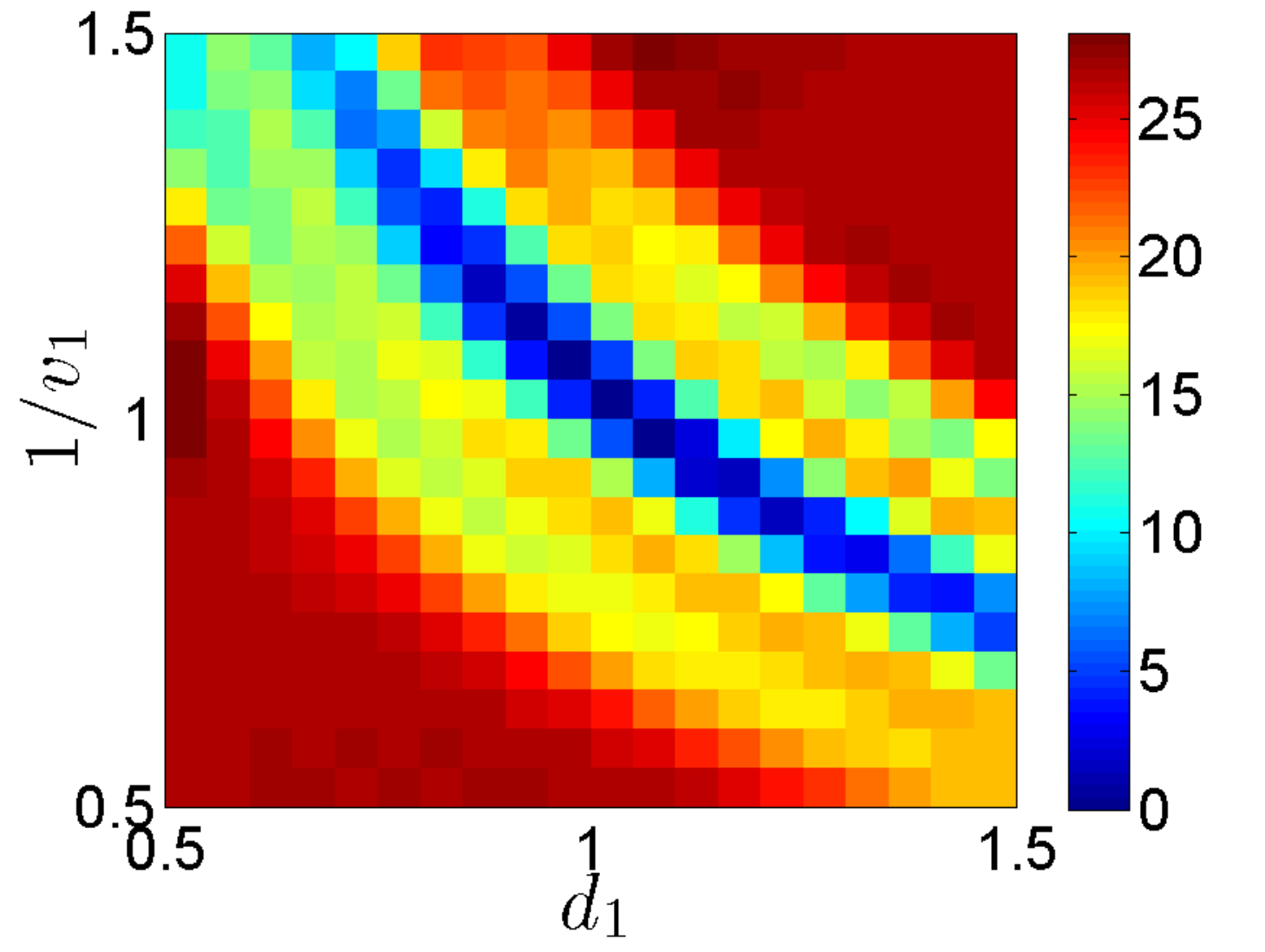}\label{fig:L2_d1v1_surf}}
\subfigure[]{\includegraphics[width=0.45\textwidth]{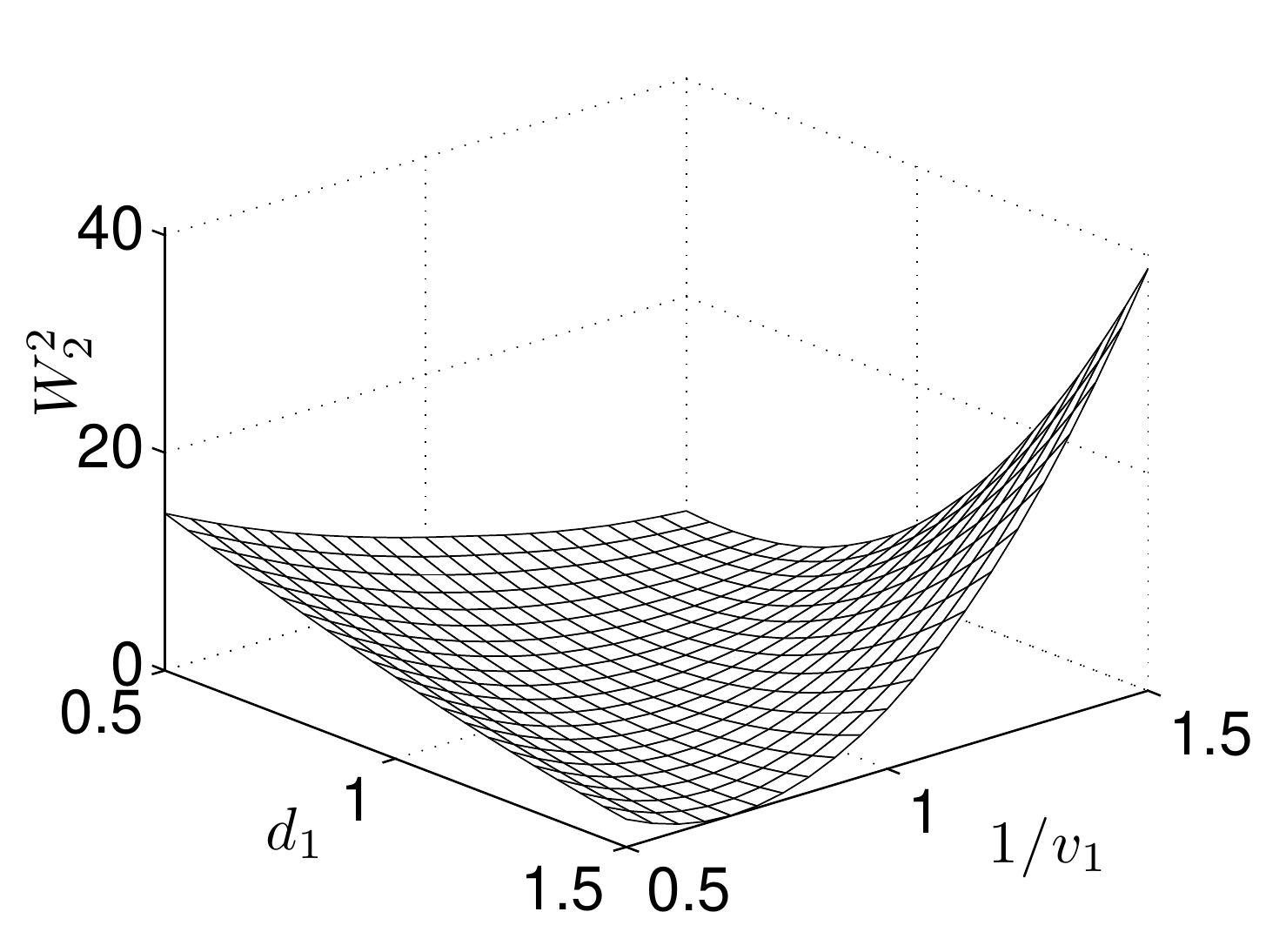}\label{fig:W2_d1v1_mesh}}
\subfigure[]{\includegraphics[width=0.45\textwidth]{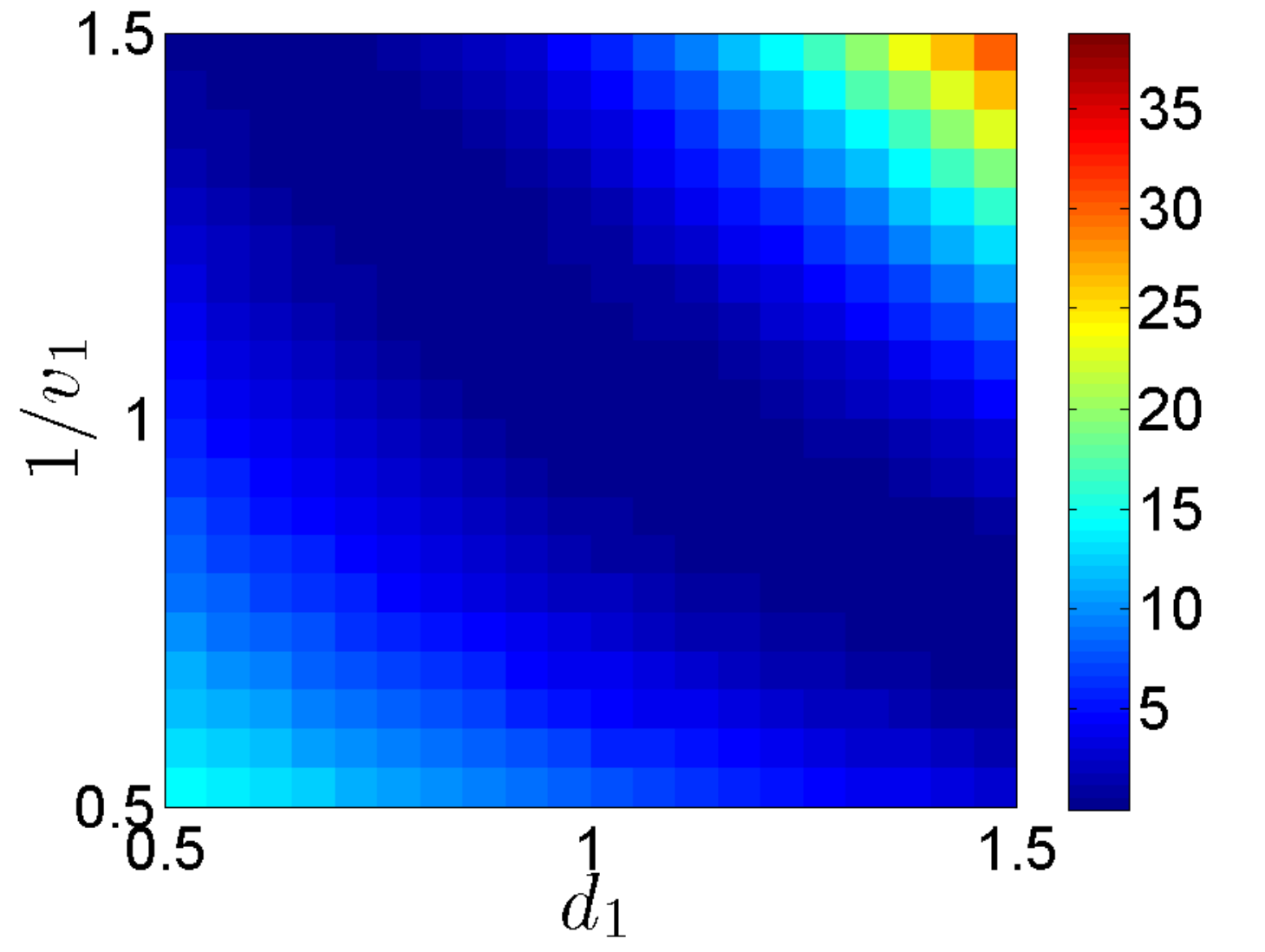}\label{fig:W2_d1v1_surf}}
\subfigure[]{\includegraphics[width=0.45\textwidth]{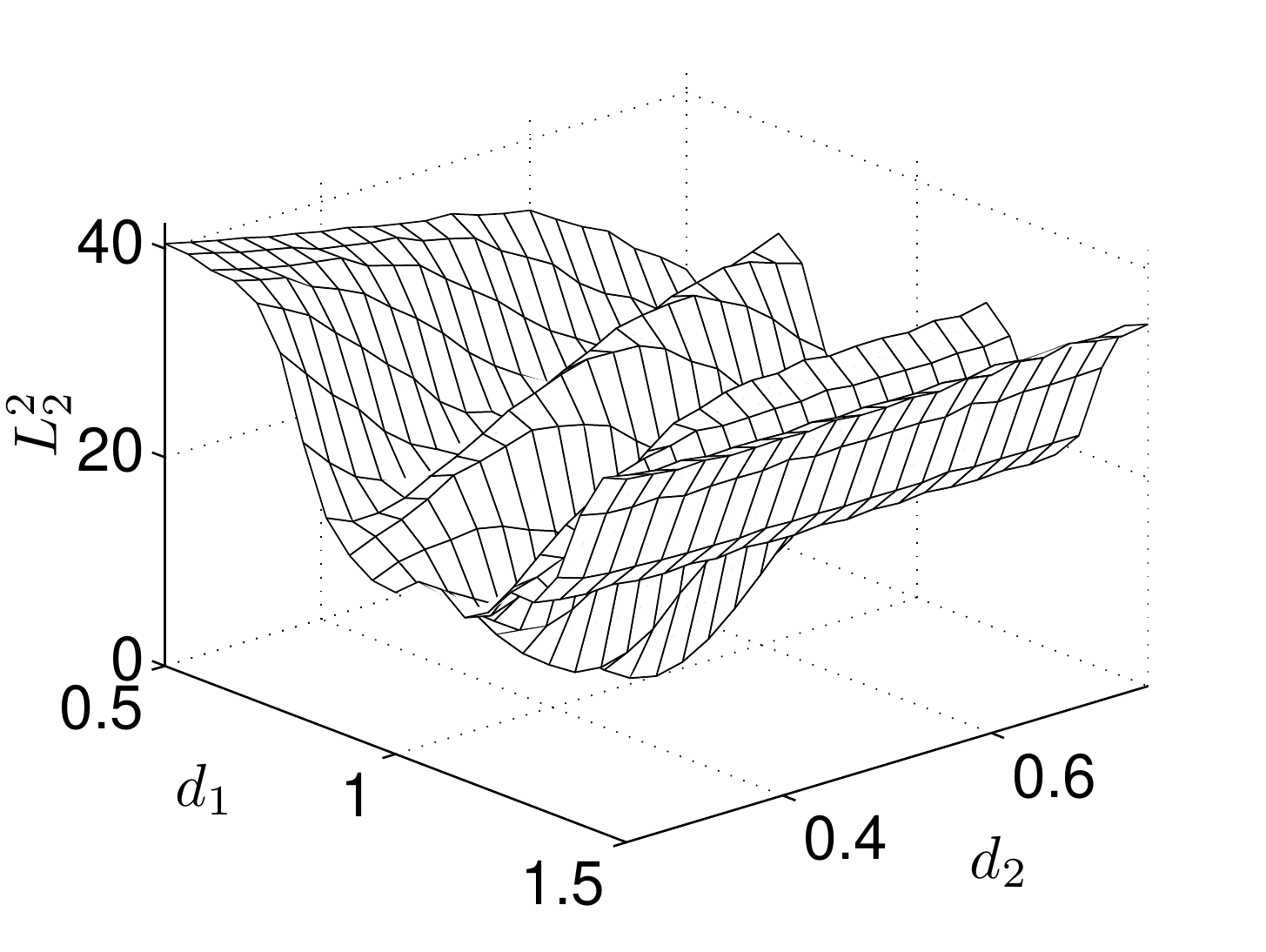}\label{fig:L2_d1d2_mesh}}
\subfigure[]{\includegraphics[width=0.45\textwidth]{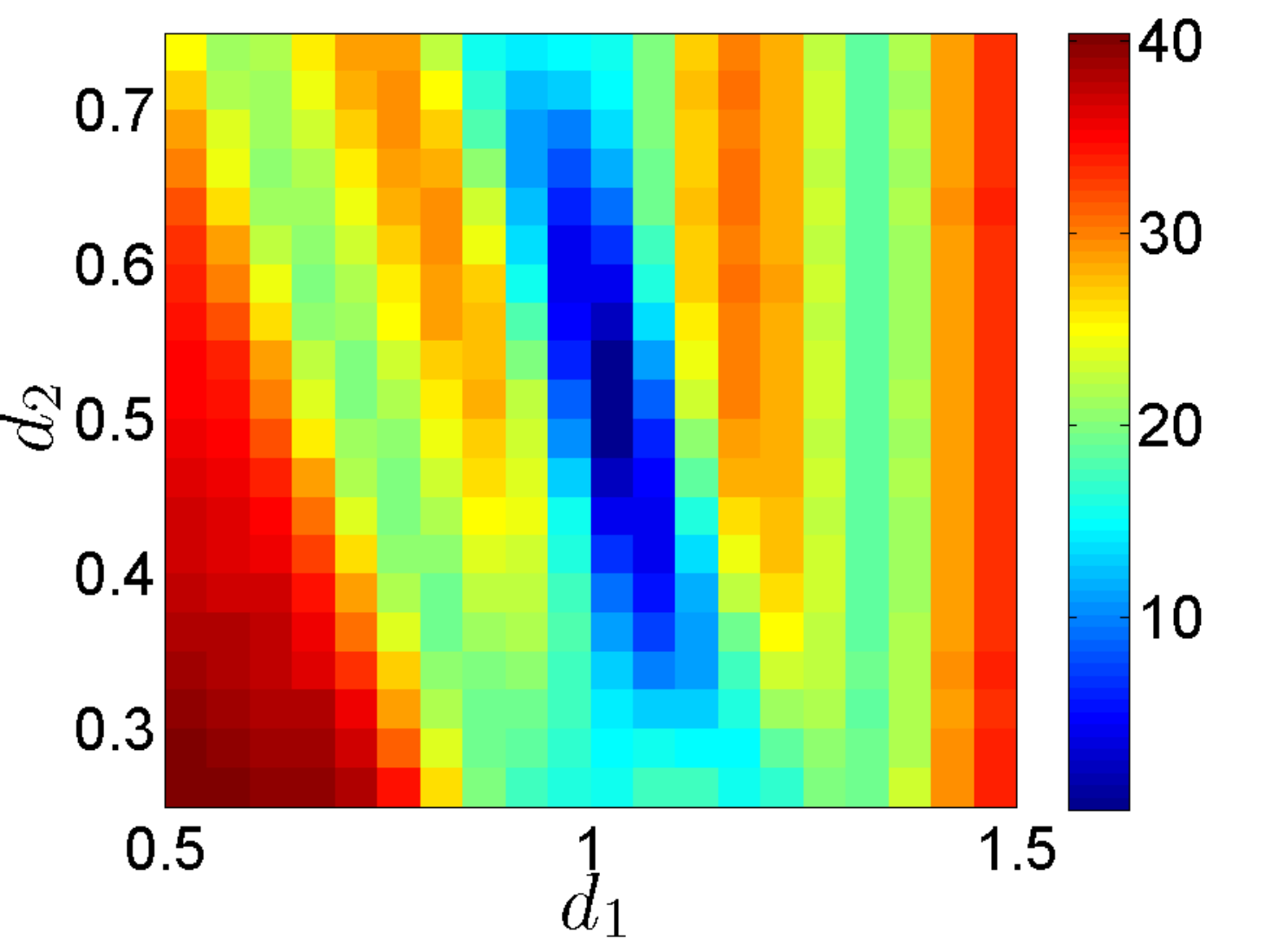}\label{fig:L2_d1d2_surf}}
\subfigure[]{\includegraphics[width=0.45\textwidth]{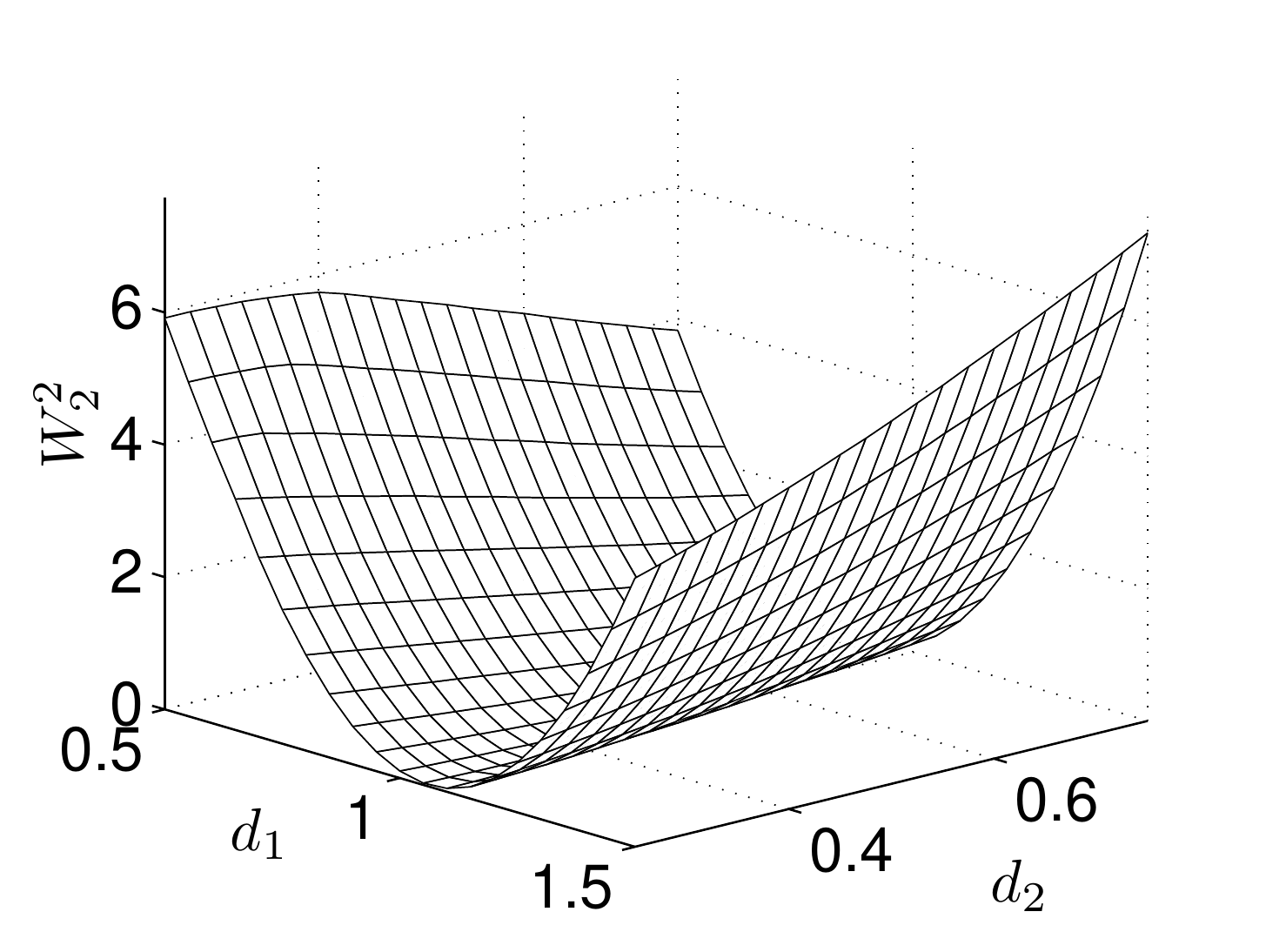}\label{fig:W2_d1d2_mesh}}
\subfigure[]{\includegraphics[width=0.45\textwidth]{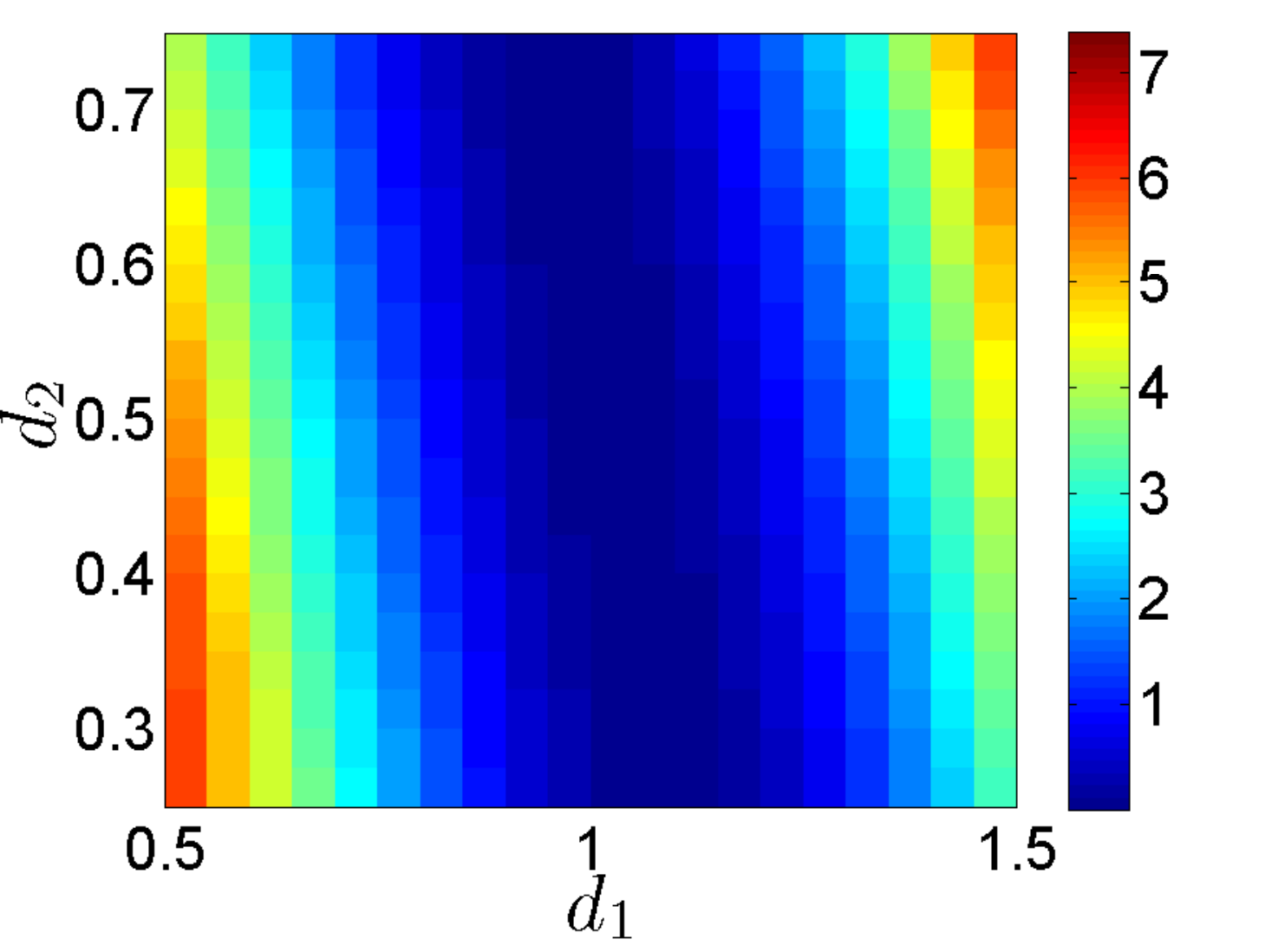}\label{fig:W2_d1d2_surf}}
\caption{Distances between profiles in Figure~\ref{fig:layer}.  Cross-sections of \subref{fig:L2_d1v1_mesh},\subref{fig:L2_d1v1_surf}~$L_2^2$ and  \subref{fig:W2_d1v1_mesh},\subref{fig:W2_d1v1_surf}~$W_2^2$ for $d_2=0.5$, $v_2=1.5$.  Cross-sections of \subref{fig:L2_d1d2_mesh},\subref{fig:L2_d1d2_surf}~$L_2^2$ and  \subref{fig:W2_d1d2_mesh},\subref{fig:W2_d1d2_surf}~$W_2^2$ for $v_1=1$, $v_2=1.5$.}
\label{fig:cross_sections}
\end{center}
\end{figure}

A second application we have in view is seismic registration.  With this in mind, we consider the (scaled) displacement vectors $\nabla u(x)-x$, which are pictured in Figure~\ref{fig:map}.  This figure indicates that the two components of the initial distribution are being transported towards the two corresponding components of the target distribution. (The small amount of artificial transport due to the preprocessing is truncated to zero).  The vector $\nabla u(x)-x$ then gives the connection between $f$ and $g$ and $\det(D^2u)$ can be used to measure the registered amplitude difference.

\subsection{Noise}
The $W_2$ metric is highly robust to noise.  The difference between a noisy signal 
\[ f(x) + h(x)>0, \quad \E[h(x)]=0 \]
and the clean signal $f(x)>0$ is typically minimal owing to the strong cancellation between nearby positive and negative values of $h(x)$.  The $L_2$ difference $\|h\|_{L_2}$ is typically substantially larger.

\begin{figure}[htdp]
\begin{center}
\subfigure[]{\includegraphics[width=0.75\textwidth]{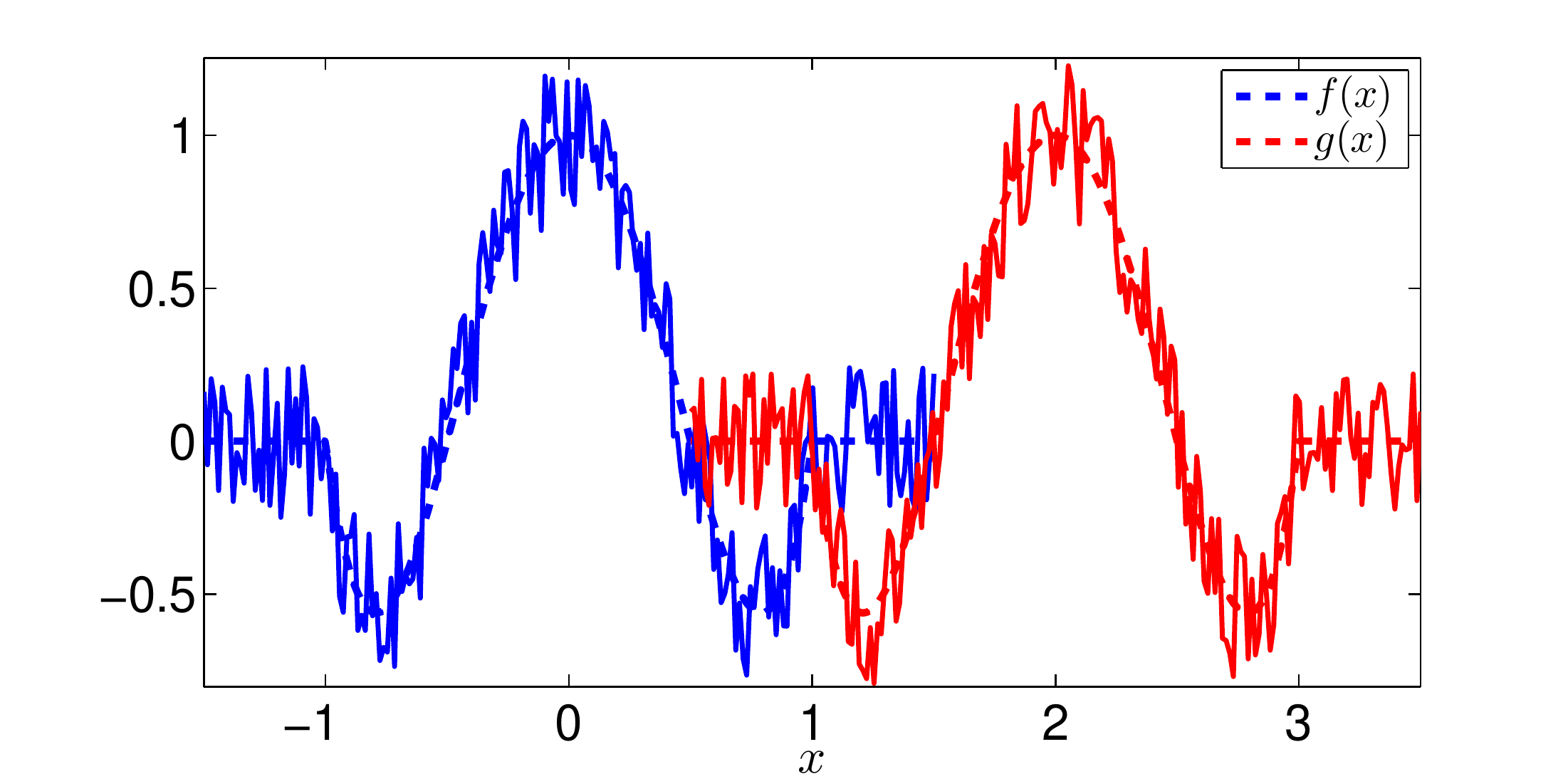}\label{fig:wavelet_prof_noisy}}\\
\subfigure[]{\includegraphics[width=0.45\textwidth]{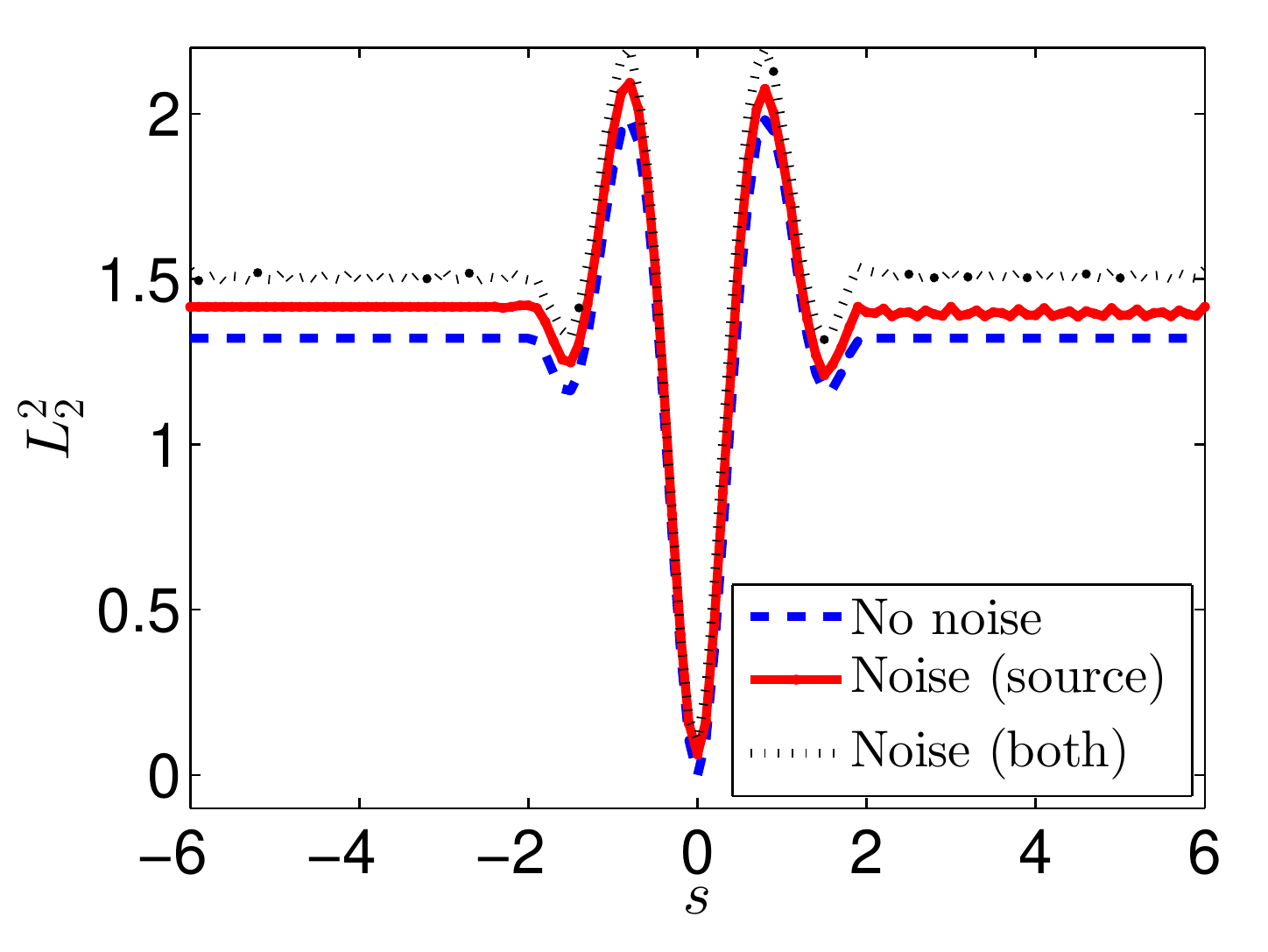}\label{fig:wavelet1d_L2_noisy}}
\subfigure[]{\includegraphics[width=0.45\textwidth]{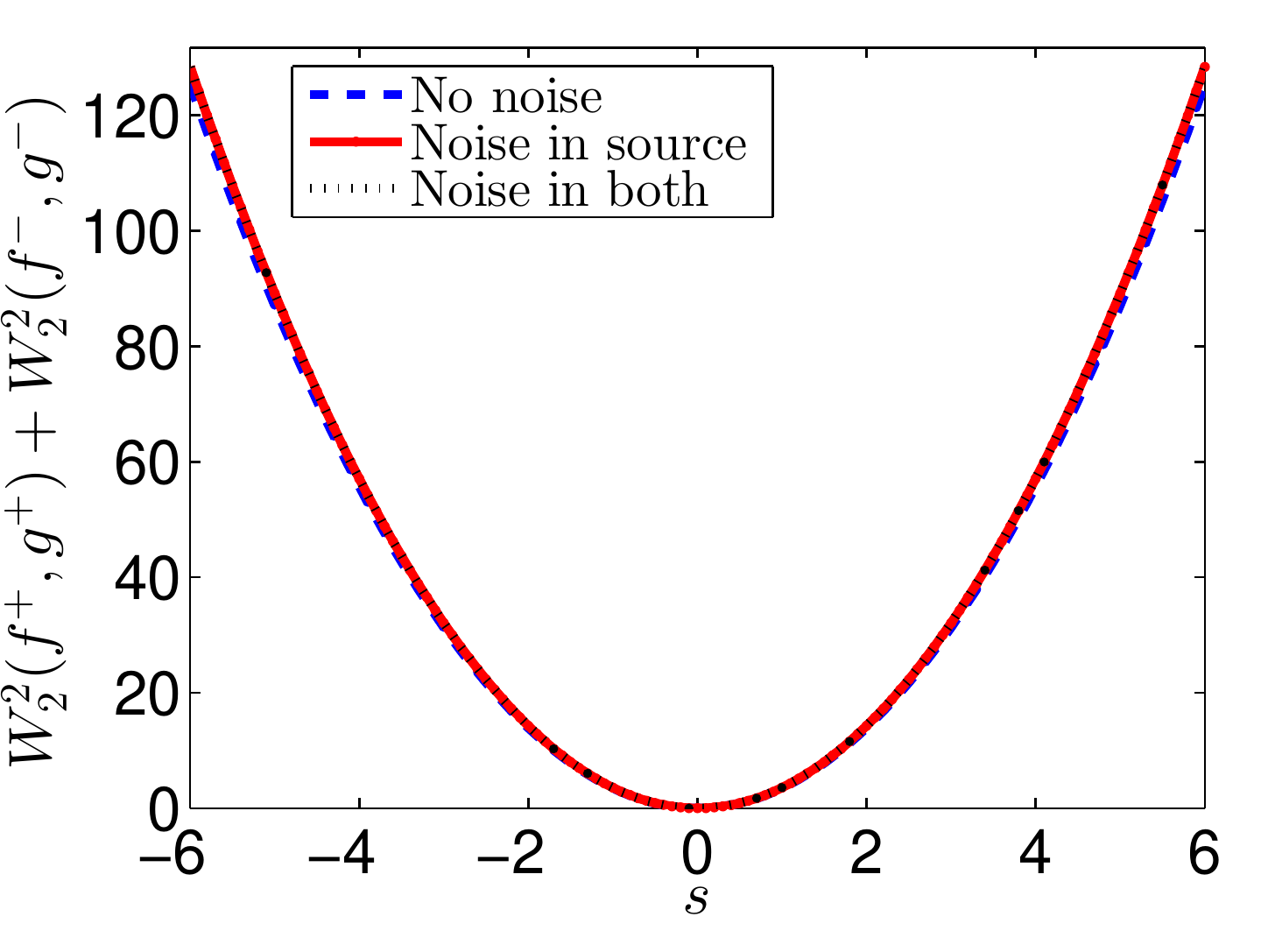}\label{fig:wavelet1d_W2_pm_noisy}}
\subfigure[]{\includegraphics[width=0.45\textwidth]{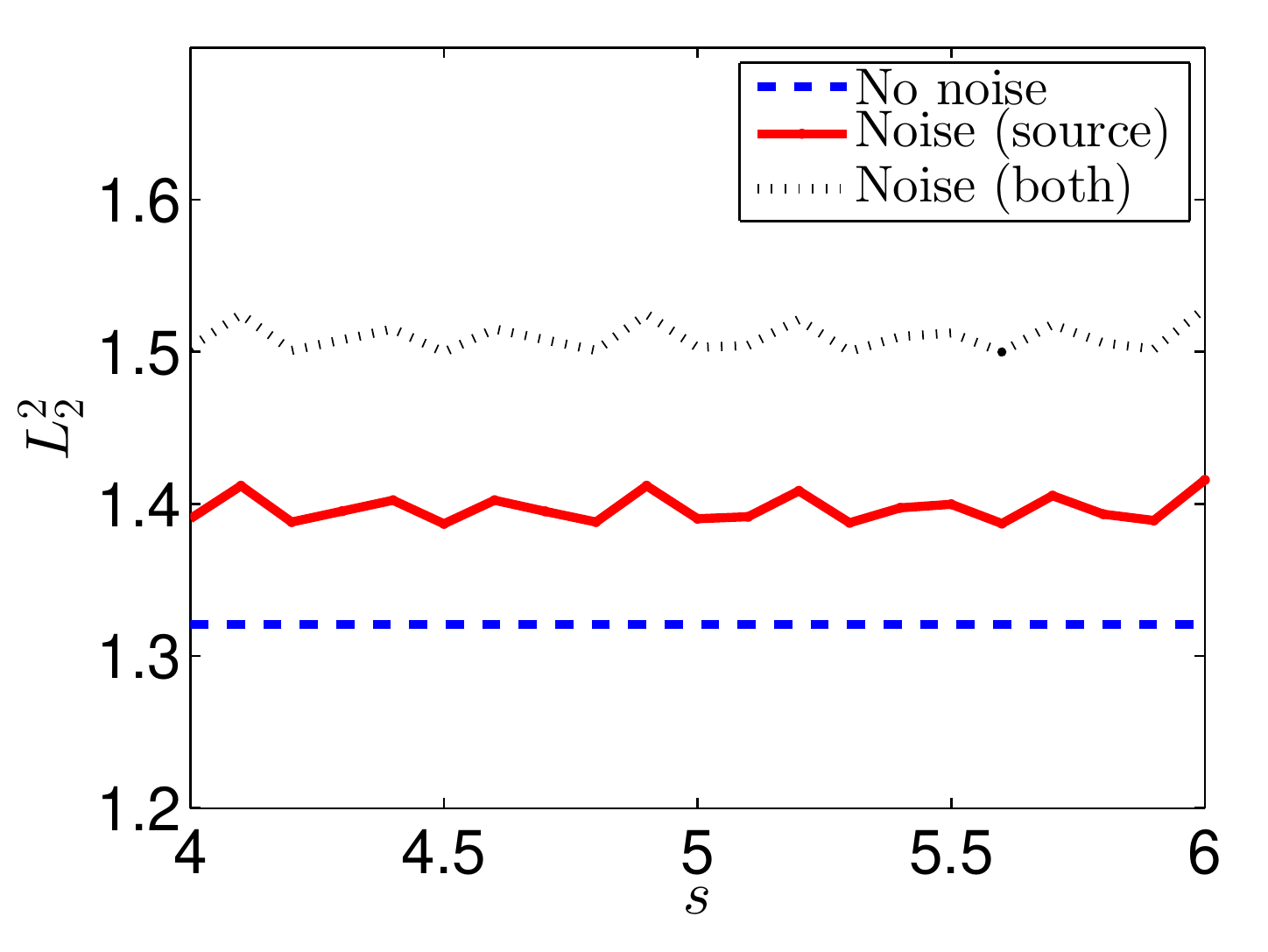}\label{fig:zoom1d_L2}}
\subfigure[]{\includegraphics[width=0.45\textwidth]{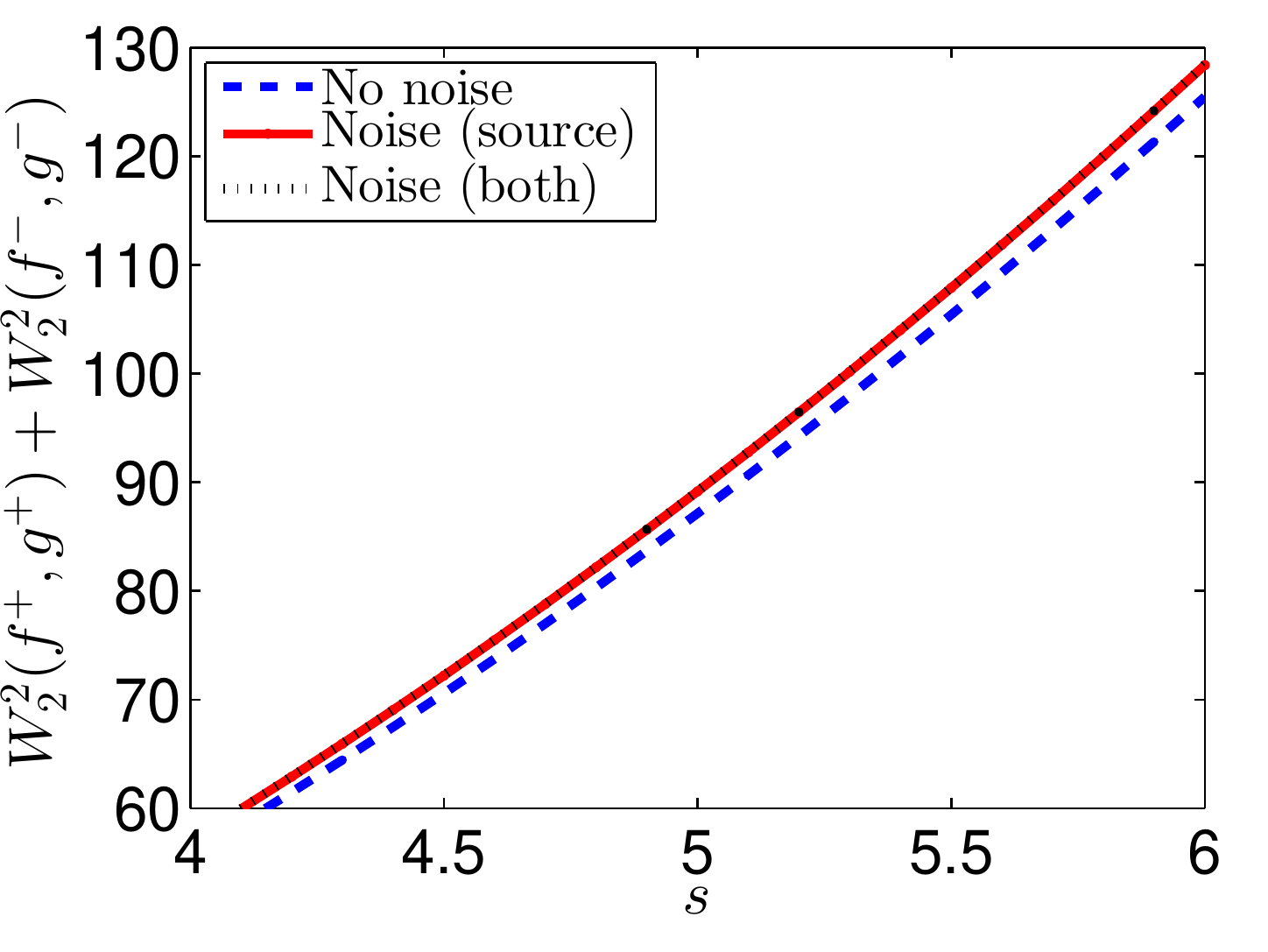}\label{fig:zoom1d_W2}}
\caption{\subref{fig:wavelet_prof_noisy}~Sample noisy profiles $f(x)$ and $g(x)$.  The distances between $f(x)$ and $g(x)$ measured by \subref{fig:wavelet1d_L2_noisy}~$L_2^2(f,g)$, and \subref{fig:wavelet1d_W2_pm_noisy}~$W_2^2(f^+,g^+) + W_2^2(f^-,g^-)$. Enlarged view of a portion of the \subref{fig:zoom1d_L2}~$L_2^2(f,g)$, and \subref{fig:zoom1d_W2}~$W_2^2(f^+,g^+) + W_2^2(f^-,g^-)$ distances.}
\label{fig:wavelet_noisy}
\end{center}
\end{figure}

To demonstrate the insensitivity to noise, we repeat the computation of the $W_2^2(f(x),f(x-s))$ distance for the wavelet profiles in Figure~\ref{fig:wavelet}.  However, this time we add uniform random noise into either the source distribution $f$ or both distributions $f, g$.  While the noise has a clear effect on the computed values of the $L_2^2$ difference, the $W_2^2$ distance computed between the noisy profiles is nearly indistinguishable from the original setting; see Figure~\ref{fig:wavelet_noisy}.  For clarity, we restrict this presentation to one dimension, but similar results are observed when we introduce noise into two-dimensional distributions.

\section{Conclusions}
We have introduced the Wasserstein metric $W_2^2$ as a measure of fidelity or misfit in seismology.  It can be seen as incorporating the most desirable properties from both the travel time difference and the $L_2^2$ difference.  We exploit recent progress in the theory of optimal transport and in fast, robust numerical methods for the \MA equation.  We further present solutions to specific challenges coming from seismic signals.  A set of simple numerical examples illustrates the advantages of this approach for potential application to full waveform inversion and registration.  Our \MA based techniques are easily generalised to higher dimensions.
\bibliographystyle{plain}
\bibliography{WassBib}

\end{document}